# Energy absorbency and phase stability during NaCl solution icing

Yanjun Shen[1,*], Xin Wei[1], Yongzhi Wang[2], Lei Li[3], Yongli Huang[4], Chang Q Sun[5,*]

Abstract

Deeper insight into the functionality of salt solute on tuning the energy absorbency R and the freezing temperature $T_N$ during phase transition is pivotal to many subject areas. Here we show that NaCl solvation turns the $f_S = n_hC$ portion molecules into the hydrating supersolid phase by ionic polarization and leaves the rest $f_O = 1 - n_hC$ portion ordinary. The C is the solute molar concentration and $n_h$ the number of molecules saturating the hydration per pair of $Na^+ + Cl^-$ solute. Polarization shortens and stiffens the H–O bond and does the O:H nonbond contrastingly in the supersolid. Water absorbs energy by H–O cooling contraction in the quasisolid phase during the process of Liquid–Quasisolid–Ice transition. The solution R drops with the $f_O$ loss till zero corresponding to C = 0.1 and $n_h$ = 10 that saturates the solvation per solute at least. The polarization-weakening of the O:H nonbonds lowers the $T_N$ < 253 K of the supersolid phase.

[1] College of Geology and Environment, Xi'an University of Science and Technology, Xi'an 710054, China (shenyj@xust.edu.cn; 19204053043@stu.xust.edu.cn)
[2] Department of Geotechnical Engineering, Tongji University, Shanghai 200092, China (yongzhi_wang@tongji.edu.cn)
[3]EBEAM, School of Materials Science and Engineering, Yangtze Normal University, Chongqing 408100, China (Lilei@yznu.edu.cn)
[4] Key Laboratory of Low-dimensional Materials and Application Technology and School of Materials Science and Engineering, Xiangtan University, Xiangtan 411105, China (Huangyongli@xtu.edu.cn)
[5] School of EEE, Nanyang Technological University, Singapore 639798, Singapore (ecqsun@ntu.edu.sg; corresponding author)

Content entry

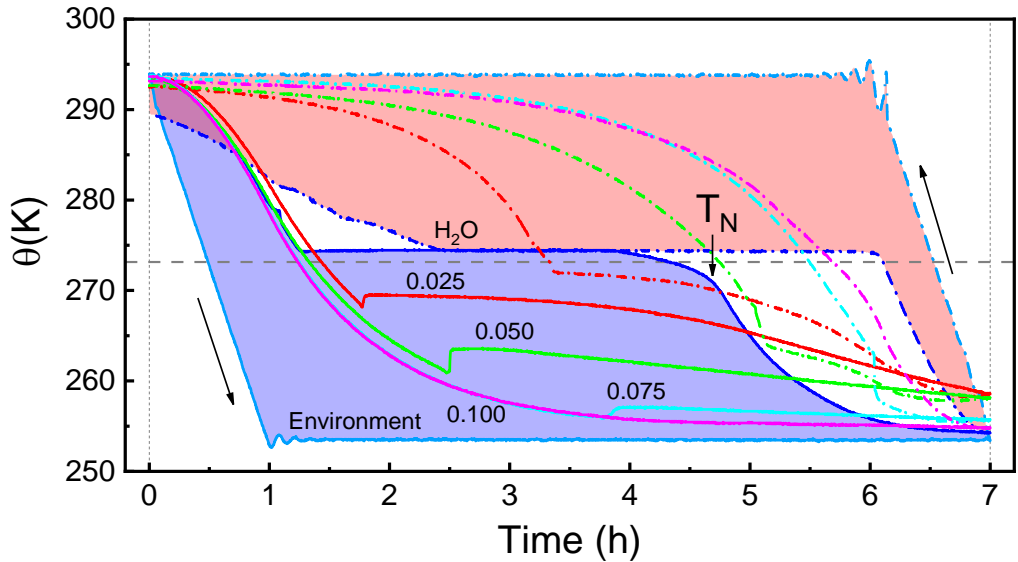

**Illustration**: Freezing-thawing loops for the concentrated NaCl solutions revealed the following:

1) H–O bond cooling contraction in the QS phase determines the fraction $f_O = 1 - n_hC$ of the residual ordinary water dependency of energy absorption ($E_{ab} \propto$ area between the cooling $\theta(C, t)$ within the QS phase and the extrapolation of that for the liquid).

2) O:H weakening in the hydrating supersolid dictates the $n_hC$ fraction dependence of $T_N$ depression.

3) At saturation of hydration, $n_hC = 1.0$ or $n_h = 10$, neither energy absorption nor $T_N \leq 253$ K is resolved.

4) Solution reduced time constant $\tau$ for the exponential extrapolation of the Liquid $\theta_{Liq}(C, t)$ further evidences the high thermal diffusivity of the supersolid phase.

5) The polarization deformed O:H–O bond possesses low deformability.

Highlights

- Ionic hydration creates the supersolid phase by length-/shortening the O:H/H–O.
- H–O bond cooling contraction absorbs energy in the quasisolid phase.
- Supersolid O:H weakening depresses the freezing point to $T_N \leq 253$ K.
- Solvation is fully saturated at $C = 0.1$ with 10 $H_2O$ molecules per solute.
- The low deformability of the supersolid hinders H–O from absorbing energy.

# 1    Introduction

Salt hydration has amazing effect on the functionality of the solution on protein dissolution and freezing temperature depression, which has profound impact to DNA engineering [1], ion channel activating and deactivating [2], and cryopreservation of living cells for disease curing and health caring [3, 4]. Salt solvation not only lowers the critical temperature $T_C$ of ice melting but also raise the temperature for ice-liquid transition. For instances, spreading salt on the snowed roads could melt ice to avoid traffic accidents; Ionic hydration could form columnar and stiffener volumes that are sufficiently strong to enlarges the graphene-oxide layer separation from 0.33 up to 1.5 nm [5, 6].

The cycling of freezing-thawing of hydrated soils and minerals are important to agricultural productivity and construction engineering safety, because the volume cooling expansion and its reverse of water molecules offers a large impact on the structure stability and properties of both water and the host substance [7, 8]. For instances, mineral rock erosion by repeated freezing-thawing is related to the safety of railway construction in the mountain areas. Rain water penetrates into pores of the substance and expands into ice in the Winter [9]. Volume expansion exerts force nearby to crack the rocks. Thawing of ice leaves the damage behind. The freezing-thawing of soil is essential to keep moisture, nutrition, and fertilizing, which is beneficial for the plant growth [10].

Therefore, comprehension of factors dictating the performance of salt solutions in terms of their bonding and electronic dynamics is pivotal to deep engineering of water and solutions to control their reaction, transition, transport dynamics and their macroscopic properties. Molecular scale understanding of salt hydration dynamics particularly in the energetic-spatial-temporal domains and its consequence on the performance of water molecules in the hydration shells remains a great challenge. Debating on the origin of the functionality of salt solvation are continuing with possible mechanisms of interaction lengths, ionic specificity [11, 12], ion-skin induction [13], quantum dispersion [14], structural order and disorder making [15-19], etc., since 1888 when Franz Hofmeister [20] firstly recognized the series of salt solubility of protein and the solution surface stress modulation.

Using Raman spectroscopy, Li et al [21] resolved the H–O stretching vibration ($\omega_H$) ranges from 3051 to 3628 cm$^{-1}$ and assigned the higher wavenumbers to water molecules with fewer than four neighbors and the lower ones to those with four ice-like hydrogen bonds. At 293 K, solvation of Cl$^-$, Br$^-$, and I$^-$ results in the same spectral features of liquid heating that shifts the $\omega_H$ band up. Hydrogen bond breaks up to form more free molecules by heating to the boiling point, which support the framework of water structure breaking or molecular separating. Conversely, Smith, Saykally, and Geissler [22] related the Raman $\omega_H$ blueshift to the degree of hydrogen bonding, instead of structure bond breaking. Based on measurements of potassium halide solutions, they contended that the spectral difference between the salt solutions and the deionized water at heating arises primarily from the electric fields beyond the first hydration shell rather than from rearrangement of the hydrogen bonds by solvation. However, it remains inspiring how salt solvation modulates the freezing point $T_N$ and what the amount and manner of energy exchange are between the solution and the environment during freezing.

This communication is focused on the mechanism behind the freezing temperature $T_N(C)$ depression and the energy absorbency $R(C)$ of NaCl solution undergoing Liquid-Quasiliquid-Ice transition, based on the framework of O:H–O bond cooperativity and polarizability (HBCP) [4, 23]. Optical imaging revealed the mode of ice and the $\theta(C, t)$ profiling revealed the quantitative $T_N$ depression and estimation of the energy absorbency during cooling from 293 to 253 K. Consistency in the quantum computations, phonon spectroscopic detection, and experimental observations confirmed that the cohesive energy of the hydrating O:H nonbond intrinsically dictates the $T_N$. Salt solvation turns the quasisolid into the supersolid. The cooling stiffening of the H–O bond in both quasisolid and supersolid phases absorbs energy. The $T_N$ drops inversely with the total number of the polarization-weakened O:H nonbonds. The R drops with solute concentration because of the low deformability of the polarization-stiffened H–O bond in the supersolid hydration cells.

## 2    Principle

First, we examined the O:H–O bond segmental length cooperativity using the *ab initio* force field calculations [24]. Fig 1 insets [25] illustrate the HBCP. The O—O Coulomb repulsion integrates the intermolecular (O:H) and intramolecular (H-O) interactions to make the coupled O:H–O bond that relax its segmental length in an unprecedented manner: both O anions dislocate in the same direction by different amounts with respect to the proton H$^+$ being taken as the coordination origin [23]. The O:H cohesive energy in the 0.10 ~ 0.22 eV range is below 5% of the H–O bond energy of 4.0 ~5.1 eV.

Second, the cohesive energy $E_x$ and the vibration frequency $\omega_x$ of each segment defines uniquely its specific heat $\eta_x(\theta/\Theta_{Dx})$ with $\Theta_{Dx}$ being the Debye temperature [25], according to Einstein relation $\Theta_{Dx} \propto \omega_x$ and to the fact that the integral of the specific heat equals the $E_x$. The subscript x = H and L represents for the H–O and the O:H segment, respectively. Interplaying the $\eta_x$ curves of the O:H–O bond defines the Vapor, Liquid, QS, ice I, and ice XI phases and their boundaries, and the density evolution of water ice in the full temperature range. The unprecedent QS phase possesses the negative thermal expansibility (NTE) and bounded by temperature of extreme densities at $T_N$ (0.92 g·cm$^{-3}$) and nearby the melting point $T_m$ (1.0 g·cm$^{-3}$). In the QS phase, cooling shortens the H–O more than the O:H elongation, resulting volume expansion and ice floating. In the Liquid and ice-I phases, O—O cooling contraction takes place by shortening the O:H and lengthening the H–O slightly. Under any circumstance, H–O contraction absorbs energy, and it emits energy at inverse [26].

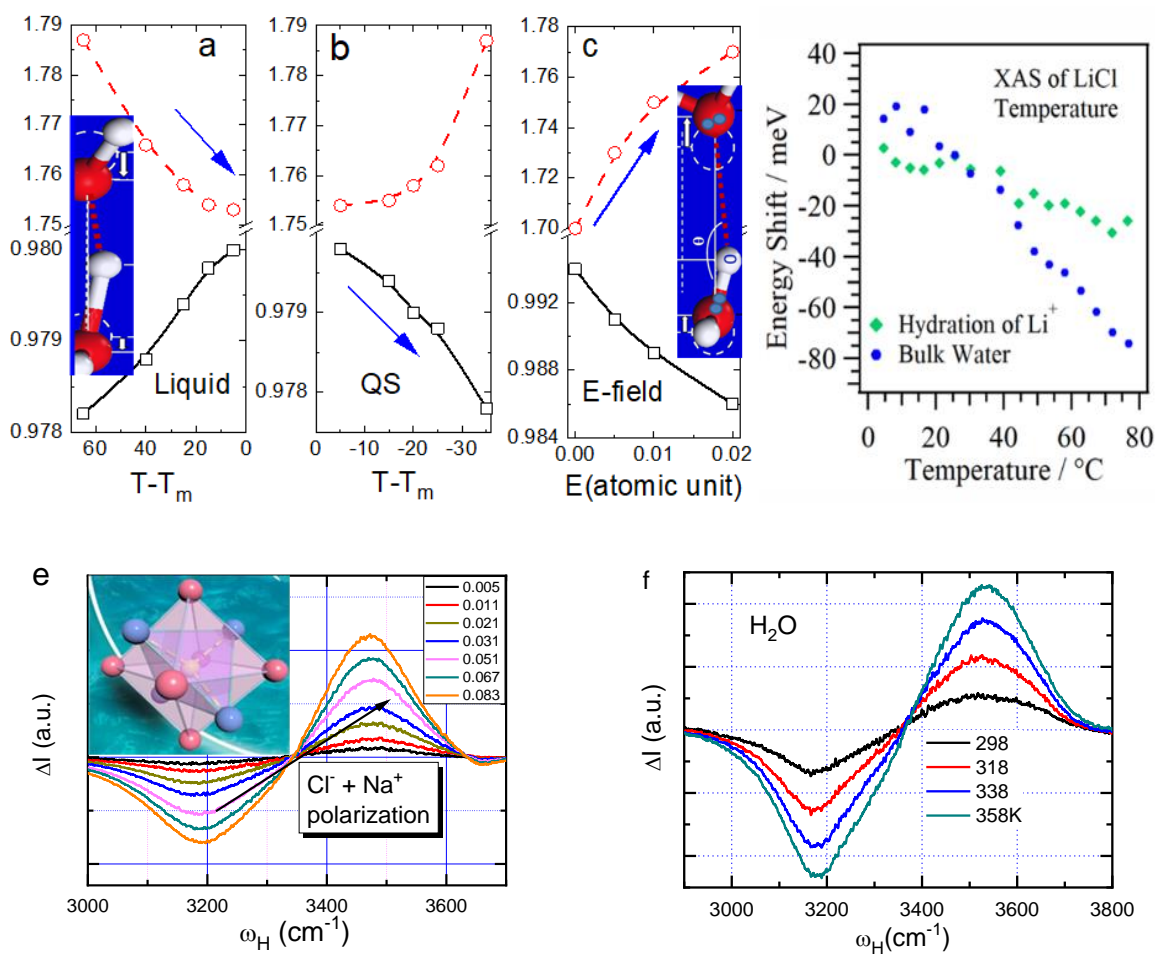

Fig 1 O:H–O bond cooperative relaxation. (a) Liquid cooling shortens the $d_L$ and lengthens the $d_H$ (inset a), while (b) QS cooling and (c) electrification (in atomic unit of 51.42 eV/Å) do the contrast (inset c). (d) The slower pre-edge shift of the XAS indicates the low deformability of the hydrating $d_H$ in the solution [27]. (e) The $\omega_H$ DPS of concentrated NaCl solutions transiting from the ordinary water at 3200 to the hydration cells of 3430 cm$^{-1}$. Heating liquid water transits (c) the $\omega_H$ from 3200 to 3500 cm$^{-1}$ or above. Inset e shows the ionic ($\pm$:4H$_2$O:6H$_2$O) hydration cell in the solution (Reprinted with permission from [4]).

Next, molecular undercoordination at sites of the skin or surrounding defects [28] or electrification by change injection such as salt solvation [4] or applying an electric field [23] shortens and stiffens the H–O bond and does the O:H the other way around, turning the ordinary water into the supersolid that is gel like, viscoelastic, mechanically and thermally more stable.

Fig 1a-c shows the cooperative relaxation of the O:H–O bond at Liquid and Quasisolid (QS) cooling and electric polarization. X-ray absorption spectroscopy (XAS) revealed that the hydrating H–O bond is thermally more stable than the unpolarized ones in the LiCl solutions [27].

The segmental cohesive energy $E_x$, critical temperature $T_C \propto E_x$, and vibration frequency $\omega_x \propto (E/d^2)_x^{1/2}$ respond cooperatively to perturbation as a function of the segmental length:

$$\left\{ \begin{array}{l} \dfrac{d\omega_x}{\omega_{x0}dq} \\ \dfrac{dT_C}{T_{C0}dq} \end{array} \right\} = -\dfrac{dd_x}{d_{x0}dq} \left\{ \begin{array}{ll} \left(1 + \dfrac{d_{x0}}{2E_{x0}}\left|\dfrac{dE_x}{dd_x}\right|\right) & (a) \\ \dfrac{d_{x0}}{E_{x0}}\left|\dfrac{dE_x}{dd_x}\right| & (b;\ x = L\ \text{or}\ H) \end{array} \right.$$

(1)

The $T_C$ depends on either the $E_L$ or the $E_H$ that is inversely proportional to a certain power of the $d_x$ [28], unless the $T_C$ is parallel to the T or P axis in the T-P phase diagram [29]. The slope $-dd_x/dq$ (q = θ, ε, z, and P) is the base for the $\omega_x$ and the $T_C$ shift and the deformability of the $d_x$. The slower decrease of the XAS O1s pre-edge shift ($\Delta$E1s) with temperature in Fig 1d indicates that the hydrating H–O bond is less deformable because the $\Delta$E1s is dominated by the $E_H$ relaxation [30]. Numerical reproduction of the $T_C$ profiles in the phase diagram confirmed that the $E_H$ dictates the $T_m$ and the $E_L$ governs the $T_N$ and $T_V$ for evaporation [29].

## 3   Phonon characteristics

The spectral peak of Raman scattering transforms all bonds vibrating in the similar frequencies in a species, irrespective of their locations, orientations, or structure phases. The spectral peak sorts out the vibration modes according to their stiffnesses of the oscillators. This Fourier transformation allows one to focus on the performance of one bond representing for its sort under stimulation. The differential phonon or spectrometrics (DPS) filters the phonon abundance (peak area integral), fluctuation (peak width), and stiffness (frequency shift) transiting from the vibration mode of regular water to that due to conditioning.

Fig 1e and f compare the $\omega_H$ Raman spectra for salt solutions and water heated from 278 to 358 K. Inset a illustrates the ionic occupancy in its solution. An ion occupies eccentrically the tetrahedral coordinated interstitial site to form the ($\pm$:4$H_2O$:6$H_2O$) hydration cell by polarizing its surrounding water molecules without bonding to water [4]. The ionic electric field polarization shortens the H–O bond and elongates the O:H nonbond, as shown in Fig 1c and further confirmed by the $\omega_H$ blue shift in Fig 1e. The former raises the $T_m$, and the latter depresses the $T_N$ and $T_V$. Indeed, salt hydration has the same effect of liquid heating on $\omega_H$ stiffening and its counterpart $\omega_H$ softening albeit different mechanisms. Liquid O:H undergoes thermal expansion and H–O contraction when heated, while salt hydration provides ionic field of polarization [22]. The consistency of the $d_x$ and $\omega_L$ relaxation verifies the formulation eq (1).

## 4   Cooling phase transition at the ambient pressure

Then, we examine the mode evolution of ice formation from deionized water and concentrated solutions deposited on Cu substrate in a 253 K bath, using optical imaging. The top two images in Fig. 2 show that the three-phase junction line and the curved ice/water interface shift up during freezing. The appearance of a spire cone on the top of the droplet indicates the end of freezing [31-34]. The sharpness of the cone does not change with the substrate temperature or the contact angle between the droplet and the substrate but it varies with the substrate materials and the nature of the solution [35].

As the concentration drops to 82 $H_2O$ per solute, the spire cone turns to be blunt. The spire cone does not appear at 55 $N_{H2O}/N_{NaCl}$ ratio. Ionic hydration polarizes and stretches the O:H–O bond by shortening and stiffening the H–O (Fig 1c and e) and lengthening and softening the O:H [36]. The electrically deformed hydrating O:H–O bond is less deformable (refer to Fig 1d) [27]. The prolonged O:H bond lowers the $T_N$ to temperature below 253 K. The observations confirm the prediction of the NTE and the viscoelasticity of the ionic hydration volume with lowered $T_N$ and low deformability, which could be advantage for the cryo-preservation of living cells.

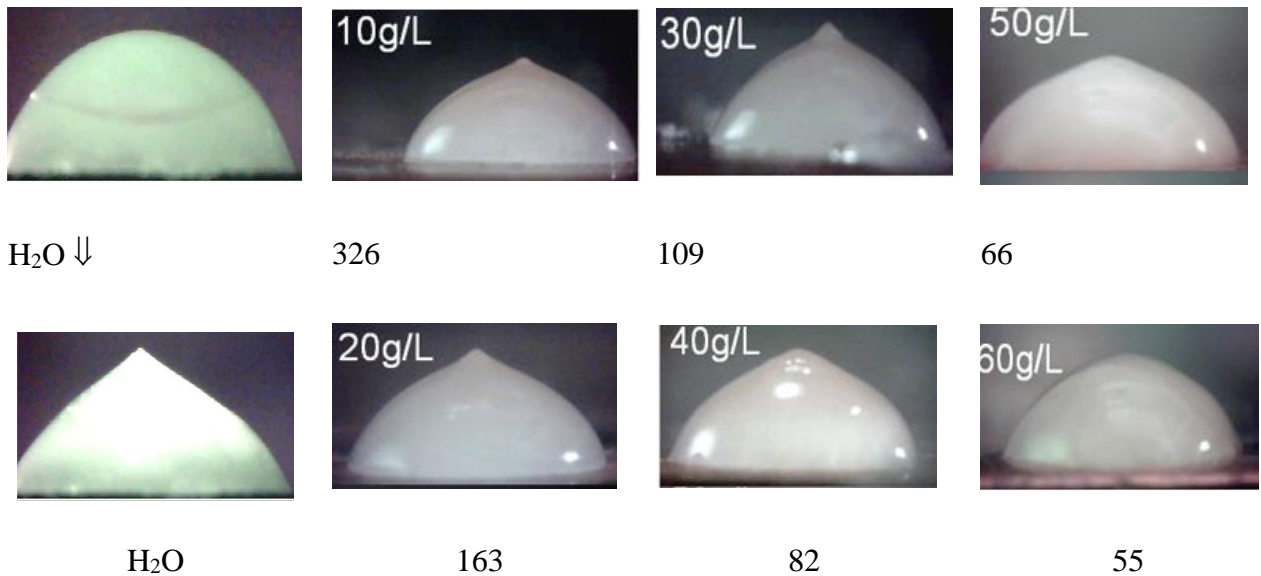

| H$_2$O ⇓ | 326 | 109 | 66 |
| H$_2$O | 163 | 82 | 55 |

Fig. 2 NaCl concentration-resolved solution icing at 253 K. When the $N_{H2O}/N_{NaCl} \leq 109$, the $T_N$ turns to be lower than 353 K with blunt spire cones on the solution droplet. The deformed supersolid O:H–O bond is thermally more stable or less deformable. Top left shows the three-phase junction line and the curved ice/water interface that move up till the end of freezing with presence of a sharp spire cone because of the volume expansion.

To verify the prediction of energy absorbency and gain quantitative information on the $T_N$ shift, experiments were conducted by cooling solutions in the 253 K bath. Samples were held in the bath at 293 K for 15 min and lowered the bath temperature to 253 K in 1h time and then kept for 6h followed by raising to 193 K in 1h and kept for 6h. The bath kept 10% constant humidity. The cooling $\theta(C, t)$ profiles provide quantitative information on the solution $T_N$ and the energy absorbency of the H–O bond during the phase transition of the concentrated NaCl solutions.

One can divide the solution into two portions, $f_S + f_O = 1$. NaCl solvation turns the $f_S = n_hC$ fraction of molecules into the supersolid hydrating molecules and leaves the residual $f_O$ ordinary. The $n_h$ is the saturation number per pair of the $Na^+ + Cl^-$ solute of the C-concentrated solution in molar ratio. In the supersolid phase, the H–O is shorter and stiffer and the O:H is longer and softer, see Fig 1c and e. The supersolid phase is thermally less deformable (Fig 1d). Furthermore, the H–O bond cooling contraction (Fig 1b) absorbs energy in the QS phase bound by $T_m$ and $T_N$. The contribution of energy loss of the O:H cooling elongation in the QS phase is negligible.

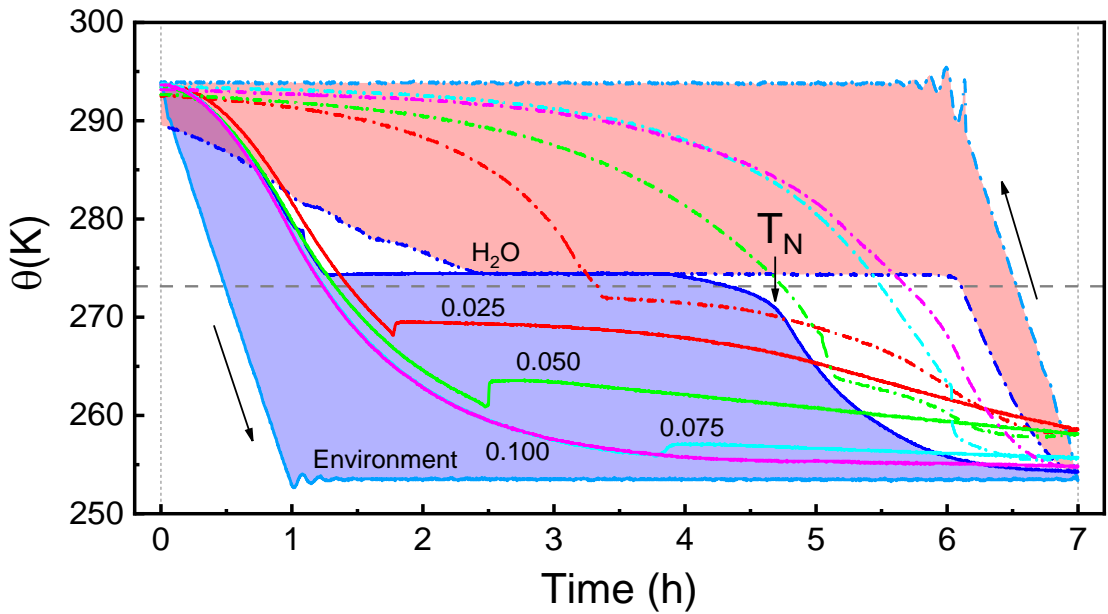

Fig. 3 The freezing-thawing $\theta(C, t)$ profiles for water and concentrated NaCl solutions show the Liquid–QS–Ice transition with indicated $T_N$ at the turning point. Deviation of the solution cooling $\theta(C, t)$ profile from the extrapolation of the liquid cooling $\theta(C, t)$ represents the energy absorption, which compensate for the energy loss during liquid cooling by H–O elongation.

The freezing $\theta(C, t)$ curves in Fig. 3 provide the following information:

1) Liquid water undergoes the Liquid–QS–Ice transition with the $T_m$ and $T_N$ as indicated in Fig. 3. The plateau between $T_m$ and $T_N$ arises from energy absorption in the QS phase through H–O cooling contraction.
2) The $T_N$ defined as the turning point of the $\theta(C, t)$ drops with the NaCl concentration till C = 0.1 with unresolvable $T_N < 253$ K. The sharp supercooled feature is suggested to be the phase precipitation. No $T_m$ could be recognized as it should be well above the $T_m$ of ordinary water [5, 6].

3) As the weakened O:H in the hydrating supersolid phase intrinsically determines the $T_N$, the detected $T_N$ is an weighted average of the $N_hC$ and $f_O$.

4) The extrapolation of the Liquid $\theta_{Liq}(C, t)$ corresponds to the energy cooing ejection by H–O bond elongation.

For the ordinary and the supersolid mixture in NaCl solution, the H–O bond energy absorption during the QS cooling, which is proportional to the area covered by the difference between the measured $\theta(C, t)$ and the extrapolation of the liquid cooling $\theta_{Liq}(C, t)$ till the $T_N(C)$ that is defined as the zero point of second differentiation of the $\theta(C, t)$ curve. The $T_N(f_S) = f_S T_N(1) + f_O T_N(0)$ is the weighted average of $T_N(1)$ for the supersolid and the $T_N(0)$ for the ordinary water. The concentration dependence of the $R(f_S)$ and $T_N(f_S)$ for the solution can be expressed as follow:

$$\begin{cases} R(f_S) & \propto \int_{T_m(f_S)}^{T_N(f_S)} C_p(f_S,\theta)\left[\theta(f_S,t)-\theta_{Liq}(f_S,t)\right]dt \quad \propto f_O \int_{T_m(0)}^{T_N(0)} E_H(0,t)dt \\ T_N(f_S)-T_N(0) & = f_S\left[T_N(1)-T_N(0)\right] \quad\quad\quad\quad\quad\quad\quad\quad\quad\quad \propto f_S\left[E_L(1)-E_L(0)\right] \end{cases}$$

(2)

Fig. 4 illustrates the estimation of energy absorbency the solution Liquid–Ice transition. We may assume the heat capacity $C_p$ as constant in the first order approximation. Table 1 lists the $T_N$, the R, and the fitting parameters as a function of the solute concentration C. The $T_N$ and R drop till C = 0.1 at which neither $T_N$ nor R could be resolved within the 253 K limit of the bath. The R is almost negligible for C = 0.073 or 13 $H_2O$ molecules hydrating per pair of $Na^+$ + $Cl^-$ solute. Therefore, the saturation number $n_h$ is within the 10-12 range. The energy absorbency drops indeed with the loss of the $f_O$ of the solution till zero, which further evidences that the shortened H–O bond in the supersolid has extremely low deformability. Furthermore, the drop of the time constant τ with the inverse of solution concentration further evidence the high thermal diffusivity of the supersolid phase [28], which fosters the Mpemba effect – warmer water cools faster [26, 37]. The time offset $T_D$ is suggested be related to the viscosity of the solution.

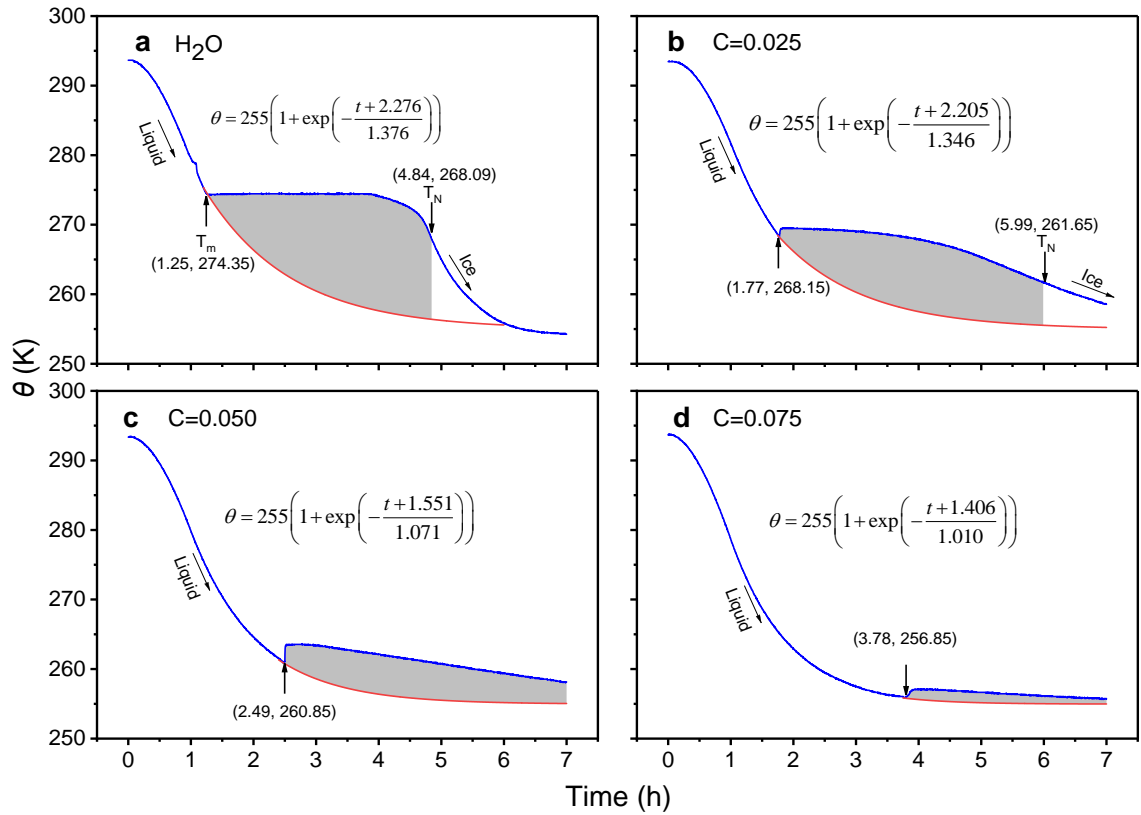

Fig. 4 Energy absorbance (shaded area) during Liquid–Ice transition of (a) deionized water and (b-d) concentrated NaCl solutions. The area is the integral of the difference between the measured θ(C, t) and the extrapolation of the liquid θ$_{Liq}$(C, t) within the quasisolid phase. No energy is absorbed at C = 0.1 (refer to appendix).

Table 1 Information of the $T_N$, R, and the fitting parameters for the Liquid–Ice transition of concentrated NaCl solutions.

| Mole Ratio | $T_N$ | $t_D$ (h) (Viscosity) | τ (h) (T-diffusivity) | Area (K·h) (E-absorbency) |
|---|---|---|---|---|
| 0 | 268 | 2.276 | 1.376 | 43.46 |
| 0.025 | 262 | 2.205 | 1.346 | 34.45 |
| 0.050 | — | 1.551 | 1.071 | 21.29 |
| 0.075 | — | 1.406 | 1.010 | 4.02 |
| 0.100 | — | 1.170 | 0.911 | 0 |

# 5 Constrained Liquid–QS–Solid transition and its application

As an extension, Fig. 5 shows the pressure constrained freezing–thawing cycling of deionized water in an artificial rock amounted with three thermal couples at the skin, center, and half-way of its radius 0.5R of a cylinder. This cycling is important to the understanding of mineral erosion and soil water fertilizing. From the HBCP point of view, the freezing-thawing is a process of pressure-constrained Liquid–QS–Ice transition and its inverse. At cooling, the thermal curve shows three regimes, correspond to Liquid, QS, and then solid Ice.

The QS phase lasts from 273 to 268 K and from 0.8 to 2.5 h to absorbs energy through H–O cooling contraction and O:H elongation that exerts forces on the rocks [38]. The inverse process emits energy by recovering the relaxed O:H–O bond with reduced constrain, which lasts for one hour at the same temperature range, as framed.

Surface molecular undercoordination and polarization lowers the $T_N$. The $T_N$ of 258 K suggests that the $T_N$ for the rock skin arises from skin supersolidity – molecular undercoordination shortens the H–O and lengthens the O:H to lower the $T_N$. The transition at the center and 0.5 R of the same plateau between 273 and 268 K and the 0.8 and 2.4 h reflects the joint temperature and inner pressure effect on the relaxation of the O:H–O bond of which the H–O cooling contraction overrides the compression elongation to absorb energy. The O:H–O bond recovery becomes easier in the process of thawing as the inner pressure of the loosened rock becomes lower.

The inner pressure limited and energy absorption with a lower $T_N$. The skin $\theta(C, t)$ profile absorbs less energy at even lower $T_N$ than the $\theta(C, t)$ at the center site of the sample, because of the involvement of the undercoordinated water molecules at the skin. The softer O:H nonbond of the supersolid skin depresses the $T_N$ and the reduced fraction of the H–O bond absorbs less energy at QS cooling. It would be promising to estimate the energy injected to the rock by integrating the difference between the $\theta(C, t)$ of water with and without the constraint.

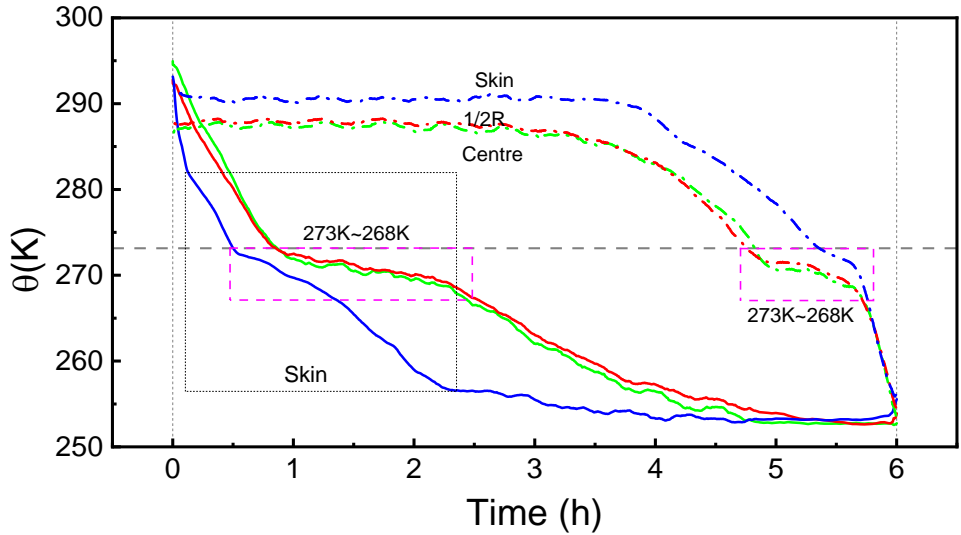

Fig. 5 Freezing-thawing loops of an artificial rock cylinder saturated with water. Thermal couples located at the center, 0.5R, and the skin of the cylinder revealed the inner pressure constrained O:H–O bond relaxation in the Liquid, QS, and Ice phases and the skin charging and molecular undercoordination effect.

## 6    Conclusion

The HBCP premise has enabled clarification of the $T_N$ shift and energy exchange during the liquid–QS–Ice transition of water and NaCl solutions and the constrained transition of water. Observations not only verify the existence of the QS phase in which the H–O absorbs energy by cooling contraction but also clarify the extraordinary elastic recoverability of the O:H–O. The understanding also furnished the entropy in the continue thermodynamics with the microscopic mechanism of bonding energy exchange through relaxation. We may conclude the following:

1)   O:H weakening dictates the $n_hC$ dependence of $T_N$ depression.
2)   H–O bond cooling contraction in the QS phase determines the $f_O$ dependency of energy absorbency.
3)   At hydration saturation, $n_hC = 1.0$ or $n_h = 10$, neither energy absorption nor $T_N \leq 253$ K is resolved.
4)   Solution reduced time constant for the exponential extrapolation of the Liquid $\theta_{Liq}(C, t)$ further evidences the high thermal diffusivity of the supersolid phase.

5) The polarization deformed O:H–O bond possesses low deformability.

Understanding may impact to the core physics and chemistry of water, ice, and aqueous solutions in terms of reaction and transition thermodynamics, which should inspire more interest and new ways of thinking about aqueous matter.

Appendix

Fitting function for the extrapolation of the Liquid freezing of NaCl solutions:

$$\theta(t,C) = A \times \left(1 + \exp\left(-\frac{t+t_D}{\tau}\right)\right)$$

For C = 0.1 Molar ratio NaCl solution freezing without $T_N$ or energy absorption:

$$\theta = 255\left(1 + \exp\left(-\frac{t+1.170}{0.911}\right)\right) \quad R^2 = 0.99984$$

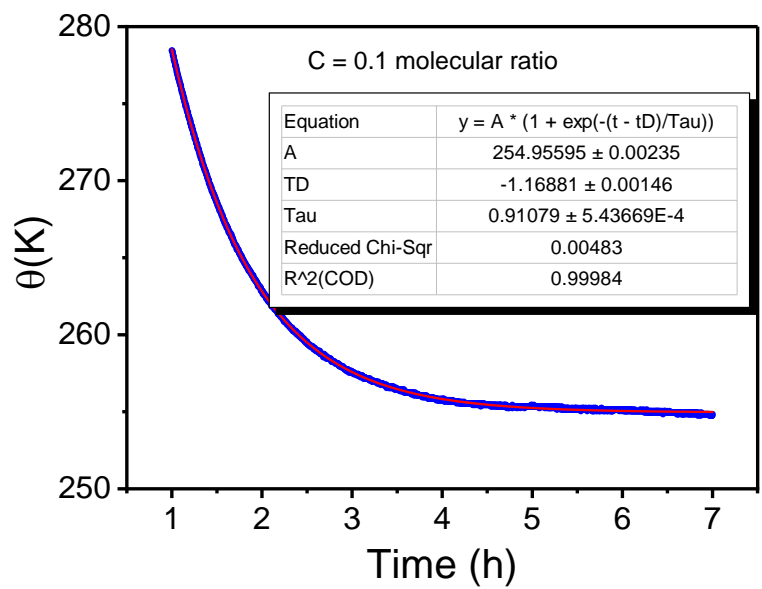

Declaration
No conflicting interest is declared.

Acknowledgement
Financial support received from National Natural Science Foundation of China (Nos. 11872052

(YL)) is acknowledged.